\documentclass[reprint,superscriptaddress,twocolumn,prl]{revtex4-2}
\usepackage{amsmath,graphicx,amssymb}
\usepackage{float}
\usepackage[usenames]{color}
\usepackage{xcolor}
\usepackage[colorlinks=true,linkcolor=blue,citecolor=red]{hyperref}%
\usepackage{siunitx}
\usepackage{physics}
\AtBeginDocument{\RenewCommandCopy\qty\SI}

\ExplSyntaxOn
\msg_redirect_name:nnn { siunitx } { physics-pkg } { none }
\ExplSyntaxOff
\DeclareSIUnit\sq{\ensuremath\Box}
\usepackage{xcolor, soul}
\sethlcolor{orange}
\AtBeginDocument{\RenewCommandCopy\qty\SI}

\newcommand\nph{n_{\rm ph}}
\newcommand\IPAT{I_{\rm PAT}}

\newcommand\hw{\hbar\omega}

\begin{document}

	\title {Efficient Microwave Photon to Electron Conversion in a\\ High Impedance Quantum Circuit}

	\author{O. Stanisavljevi\'c}
	\affiliation{Universit\'e Paris-Saclay, CNRS, Laboratoire  de  Physique  des  Solides,  91405 Orsay,  France}

	\author{J.-C. Philippe} 
	\affiliation{Universit\'e Paris-Saclay, CNRS, Laboratoire  de  Physique  des  Solides,  91405 Orsay,  France}
	
	\author{J. Gabelli} 
	\affiliation{Universit\'e Paris-Saclay, CNRS, Laboratoire  de  Physique  des  Solides,  91405 Orsay,  France} 
	
	\author{M. Aprili} 
	\affiliation{Universit\'e Paris-Saclay, CNRS, Laboratoire  de  Physique  des  Solides,  91405 Orsay,  France} 
	
	\author{J. Estève}
	\affiliation{Universit\'e Paris-Saclay, CNRS, Laboratoire  de  Physique  des  Solides,  91405 Orsay,  France}
	
	\author{J. Basset}
	\email{julien.basset@universite-paris-saclay.fr}
	\affiliation{Universit\'e Paris-Saclay, CNRS, Laboratoire  de  Physique  des  Solides,  91405 Orsay,  France}

	%\pacs{74.45.+c, 74.40.Gh}
	\date{\today}
	
	\begin{abstract} 
		We demonstrate an efficient and continuous microwave photon to electron converter with large quantum efficiency ($83\%$) and low dark current. 
		These unique properties are enabled by the use of a high kinetic inductance disordered superconductor, granular aluminium, to enhance light-matter interaction and the coupling of microwave photons to electron tunneling processes.
		As a consequence of strong coupling, we observe both linear and non-linear photon-assisted processes where 2, 3 and 4 photons are converted into a single electron at unprecedentedly low light intensities. Theoretical predictions, which require quantization of the photonic field within a quantum master equation framework, reproduce well the experimental data.  
		This experimental advancement brings the foundation for high-efficiency detection of individual microwave photons using charge-based detection techniques. 
	\end{abstract}
	\maketitle
	
	\textbf{\textit{Introduction---}}
	In the optical domain, the photoelectric effect is the method of choice to build single photon detectors covering a wide frequency spectrum with large quantum efficiency and low dark current \cite{eisaman2011}. Reducing the photon energy down to microwave frequencies ($\sim \qty{10}{\GHz}$), while fulfilling the requirements for single photon detection, is experimentally challenging because of the absence of semiconducting or superconducting materials with a sufficiently low energy gap. Consequently, efficient single microwave photon detectors rely so far on alternate strategies using the circuit quantum electrodynamics toolbox, which has its own limitations \cite{Sankar2016,Inomata2016,Besse2018,Lescanne2020,Blais2021,Albertinale2021,Wang2023}. Recently, experiments in hybrid circuits combining semiconducting quantum dots with high finesse microwave cavities have monitored photon-assisted tunneling processes using single electron charge detectors \cite{Wong2017,Cornia2021,Khan2021,Haldar2024a,Haldar2024b}. However the conversion efficiency of the incoming photons into electrons remained low. 
	
	In this letter, we demonstrate that photon-assisted quantum tunnelling of quasiparticles in a voltage-biased superconducting tunnel junction, which has been known for a long time \cite{Dayem1962,Tien1963,Tucker1985,Worsham1991,Basset2012,Tan2017,Viitanen2023}, can be pushed close to the limit where every incoming photon is converted into a tunneling electron. The impedance matching of the microwave mode to the tunnel junction is achieved by using a high impedance material, granular aluminum (grAl), which is a high kinetic inductance disordered superconducting material with interesting material properties \cite{Annunziata2010,Samkharadze2016, Grunhaupt2018,Maleeva2018,Kamenov2020,Moshe2020}. We reach 83\% quantum efficiency while keeping a low dark current ($3.4\times10^5$\,e/s) through the junction in the absence of incoming light. From a more fundamental perspective, our experiment addresses the question of photon-assisted tunneling in the regime where a quantum description of the microwave signal is mandatory to reach a quantitative understanding of the process.  Our work is complementary to previous works dealing with Cooper pairs tunneling in high impedance environments \cite{Chen2011,Leppakangas2018,Jebari2018,Albert2023,Crescini2023,Mehta2023} and extends it to incoherent quasiparticle tunneling \cite{Esteve2018,Aiello2023}. In particular, we observe high order processes where photon-assisted tunneling takes place through the absorption of 2, 3 and 4 photons, while remaining in the quantum regime, where the mean occupancy of the microwave mode is on the order of a few photons. The non-linear power dependence of the tunneling current for these different processes allows us to obtain an absolute calibration of the incoming photon flux.

\begin{figure}[tbp]
		\begin{center}
			\includegraphics[width=8.5 cm]{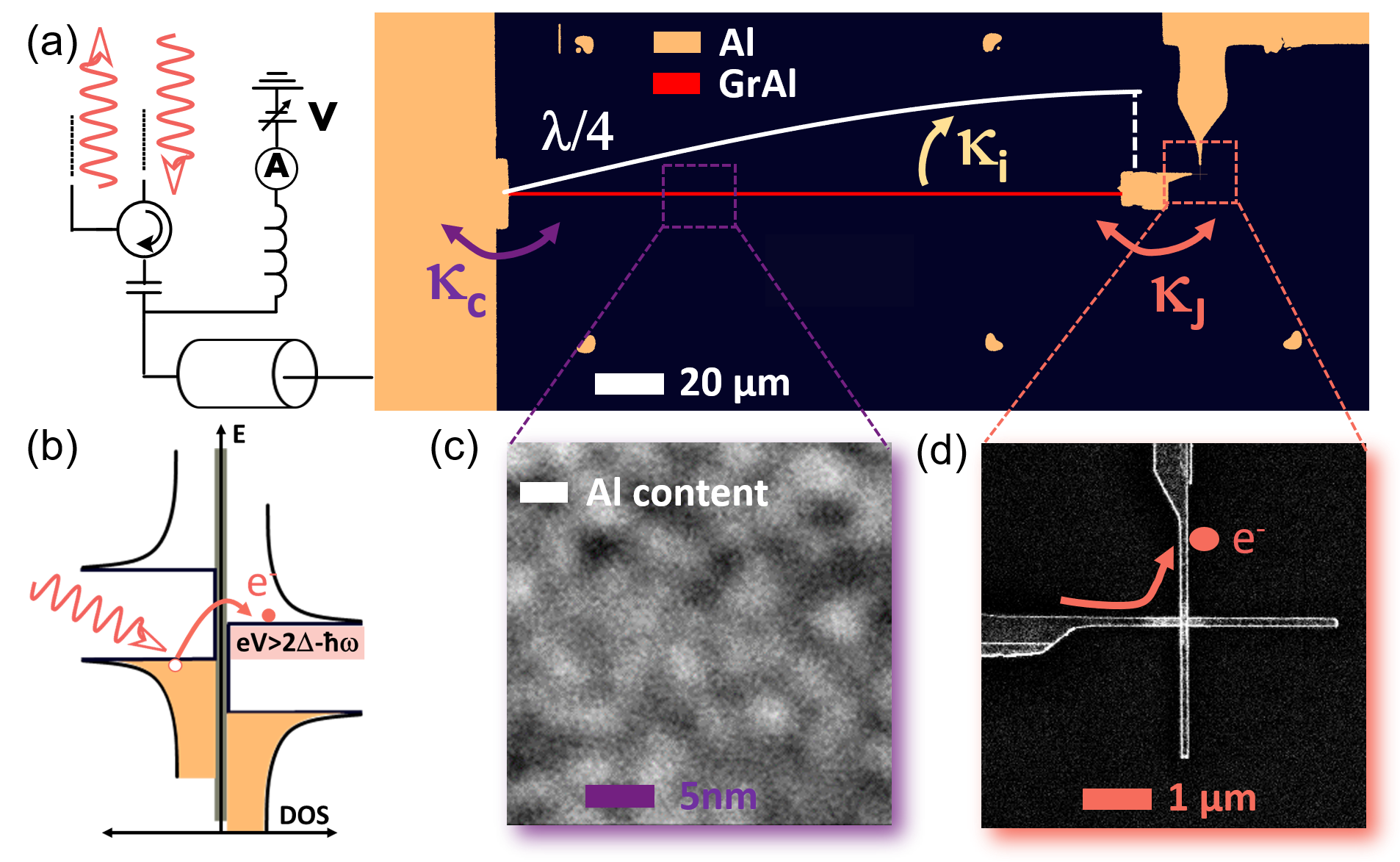}
		\end{center}
		\vspace{-0.8cm}
		\caption{\textbf{Principle of the experiment.} (a) Microscope image of the device and simplified schematic of the measurement setup. An incoming microwave photon is efficiently converted into an electron by coupling a waveguide to a high impedance resonator (red). Efficient conversion is reached when the loss rate $\kappa_j$ due to photon-assisted tunneling through the junction matches the coupling rate $\kappa_c$ to the input waveguide. The quarter wavelength resonator is made out of granular aluminum, the junction is aluminium based. (b) Energy diagram (E - vertical axis) of the displaced superconducting density of states (DOS - horizontal axis) of the voltage biased superconducting tunnel junction and a sketch of the photon-assisted tunneling of quasiparticles. The bias voltage sets a frequency band in which photons can be absorbed, while no elastic current flows through the junction. (c) Chemical content sensitive transmission electron microscope image of a granular aluminum thin film. Brighter colors mean larger aluminium content. Darker areas have a stronger oxygen content. (d) Scanning electron microscope image of the superconducting tunnel junction.}
		\label{Figure1}
		\vspace{-0.4cm}
	\end{figure}

	\textbf{\textit{Principle of the experiment---}}
	The principle of the experiment is shown in figure \ref{Figure1}. A dc-biased superconducting tunnel junction is connected to a quarter wavelength resonator made of grAl with a high characteristic impedance. The microwave photons that we aim to convert into electrons are fed to the resonator by a microstrip \qty{50}{\ohm} waveguide visible on the left of figure \ref{Figure1}(a). The photon-assisted tunneling of one quasiparticle through the absorption of one photon  from the resonator mode is energetically allowed if the bias voltage $V$ satisfies $eV > 2\Delta-\hbar \omega$, where $\Delta$ is the superconducting gap and $\omega$ the frequency of the resonator (see figure \ref{Figure1}(b)). At low temperature, and in the absence of photons, no dc current flows through the junction if $eV<2\Delta$. The working voltage range of the detector is therefore $2\Delta-\hbar \omega< eV <2\Delta$, where the dark current is low and the one photon photon-assisted tunneling process is allowed.
	The probability that an incoming photon is absorbed and converted into an electron defines the detector quantum efficiency, which reaches unity when the coupling rate $\kappa_c$ from the resonator to the waveguide matches the absorption rate $\kappa_j$ due to the photon-assisted tunneling (PAT) process. This is equivalent to an impedance matching condition. Here, we have supposed for simplicity that other internal losses are negligible. 
	
	Below the gap, the PAT rate coming from single photon absorption \cite{Devoret1990,Ingold1992,Souquet2014,Esteve2018} is given by
	\begin{equation}
		\kappa_j = \lambda^2 \, \exp(-\lambda^2) \, I(V+\hbar \omega/e)/e \label{eq.kappaj}
	\end{equation}
	where the coupling constant $\lambda=\sqrt{\pi Z_C/R_K}$ is set by the ratio of the characteristic impedance of the mode $Z_c$ to the quantum of resistance $R_K = h/e^2$ and $I(V)$ is the current voltage characteristic of the junction. In the experiment presented here, we designed the grAl resonator to reach $\lambda \sim 0.8$.
	The coefficient $\lambda^2 \rm{exp}(-\lambda^2)$ corresponds to the Franck-Condon matrix element $|\mel{0}{D_\lambda}{1}|^2$, where $D_\lambda$ is the displacement operator $\exp(i \lambda(a+a^{\dagger}))$, which translates the charge degree of freedom of the resonator mode $a$ by one electron \cite{Devoret1990}.
	Connecting the resonator to a $50\,\unit{\ohm}$ waveguide typically leads to a coupling rate $\kappa_c \sim 2\pi \times \qty{80}{\MHz}$ (see Supplemental Material~\cite{SI}). 
	
	A straightforward estimate based on equation (\ref{eq.kappaj}) indicates that a junction with a resistance of about \qty{1.5}{\Mohm} fulfills the impedance matching condition in the detector bias window~\cite{SI}. This large value of the tunnel junction resistance is highly beneficial to reduce the elastic tunneling rate and therefore the dark current, which is a key figure of merit.
	In a conventional low impedance resonator, the Franck-Condon factor $\lambda^2 \exp(-\lambda^2)$ would reduce the photon-assisted tunneling rate by approximately two orders of magnitude. The matching condition could still be reached, but with a much more transparent junction, which is detrimental in terms of dark current.
	Assuming a detector with a \qty{200}{\ohm} characteristic impedance and the same bandwidth would require a tunnel resistance about \qty{100}{\kilo\ohm}. With a Dynes parameter of \qty{10}{\nano\eV} \cite{Tan2017}, a subgap current of roughly \qty{2}{\pA} ($\approx14$\,Me/s) is then expected at the operating voltage.	
	The sample shown in figure \ref{Figure1} is fabricated via a three steps process using standard e-beam and optical lithography techniques.  
	It is then mounted in  a dilution fridge with a base temperature of \qty{20}{\milli\K} with a measurement circuit, which allows us to dc-bias the tunnel junction, feed microwaves and measure the sample microwave reflection $S_{11}$. The measured resonator has a characteristic impedance of \qty{5.1}{\kilo\ohm} ($\lambda=0.79$), and the normal state resistance of the junction, obtained by triple oxidation, is \qty{1.53}{\mega\ohm}~\cite{SI}.

\begin{figure}[btp]
		\begin{center}
			\includegraphics[width=8.5cm]{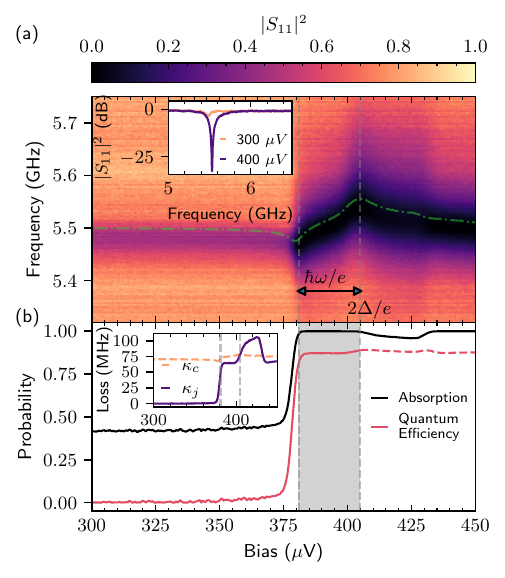}
		\end{center}
		\vspace{-0.9cm}
		\caption{\textbf{Microwave spectroscopy.} (a) Squared modulus $|S_{11}|^2$ of the microwave reflection coefficient near resonance \textit{vs} frequency and bias voltage. The resonance frequency is shown as a green dash-dotted line and undergoes Lamb shift. Inset: frequency line cuts for two significant voltages (reference and critical coupling spectra). (b) Microwave absorption and expected quantum efficiency of photon to electron conversion deduced from the spectroscopic measurements shown in (a). The quantum efficiency is meaningful only for $eV<2\Delta$, where elastic tunneling is negligible. Inset: Voltage bias dependence of the coupling and junction induced loss rates $\kappa_c$ and $\kappa_j$.}
		\label{Figure2}
		\vspace{-0.4cm}
	\end{figure}

	\textbf{\textit{Microwave spectroscopy---}}
	We show in figure \ref{Figure2}(a) a colormap of the reflection coefficient $|S_{11}|^2$ of the resonator as a function of frequency and bias voltage in the vicinity of the superconducting gap $2\Delta/e$. The power of the probe tone is sufficiently low (-142\,dBm) to keep the resonator population below 0.02 in order to be in the linear regime, where the measured spectrum is power independent.
	Below \qty{375}{\uV}, the reflection coefficient $S_{11}$ exhibits a resonance around $\omega/2\pi=\qty{5.525}{\GHz}$ with a small dissipative response (-2\,dB dip, see inset), indicating that the internal losses of the resonator are small, about $0.13 \, \kappa_c$, as desired.
	When the voltage increases and reaches $(2\Delta - \hbar \omega)/e \approx \qty{379}{\uV}$, photons are absorbed by the junction through photon-assisted electron tunneling. The reflection dip correspondingly becomes more pronounced (inset figure \ref{Figure2}(a)). Additionally, the resonance frequency shifts (dotted-dashed green line in figure \ref{Figure2}(a)) due to the coupling to the junction, illustrating Kramers-Kronig relations \cite{Aiello2023,Worsham1991,Basset2012}. 
	At a given voltage, we fit the reflection $S_{11}$ using 
	\begin{equation}
		\label{eq.S11}
		S_{11}=1-\frac{\kappa_c}{\kappa/2+\kappa_c/2+i \delta}.
	\end{equation}
	where $\delta$ is the detuning between the source and the resonance frequency of the mode $\omega$, which we fit together with $\kappa_c$ and $\kappa$. The loss rate $\kappa$ is the sum of the PAT loss rate $\kappa_j$ and other intrinsic losses with a rate $\kappa_i$. We determine $\kappa_i = 2\pi\times \qty{9.5}{\MHz}$ by taking the averaged value of $\kappa$ below $\qty{340}{\uV}$, where $\kappa_j \approx 0$. We then assume that $\kappa_i$ is voltage independent and obtain $\kappa_j=\kappa-\kappa_i$. The evolution of $\kappa_c$ and $\kappa_j$ as a function of voltage is shown in the inset of figure \ref{Figure2}(b). The coupling rate $\kappa_c$ only weakly depends on bias and is equal on average to $2\pi\times \qty{75}{\MHz}$ in the range $2\Delta-\hbar\omega< eV < 2\Delta$, where the junction loss rate $\kappa_j$ reaches $2\pi\times \qty{65}{\MHz}$ in good agreement with (\ref{eq.kappaj}) \cite{SI}. Above the gap, the elastic quasiparticle current leads to an out-of-equilibrium electronic distribution, which increases $\kappa$. When $eV$ becomes larger than $2\Delta+\hbar\omega$, $\kappa$ decreases abruptly as a consequence of photon emission by the junction into the resonator through inelastic tunneling.

	From these values, we compute the probability that an incoming photon at resonance is absorbed  $1-|S_{11}|^2=1 - (\kappa - \kappa_c)^2/(\kappa+\kappa_c)^2 $ (see figure \ref{Figure2}(b)).
	Critical coupling, which corresponds to perfect absorption ($|S_{11}|^2=0$), is reached when $\kappa_j = \kappa_c - \kappa_i$. We observe a minimal reflection below -30\,dB at the corresponding voltage (inset figure \ref{Figure2}(a)).
	We can estimate the quantum efficiency by noting that the resonator population at resonance is $\nph=4\phi \kappa_c/(\kappa_c+\kappa)^2$ where $\phi$ is the incoming photon flux, which is proportional to the incident power $P$ through $\phi=P/\hbar\omega$. The quantum efficiency is then $\kappa_j \nph /\phi = 4 \kappa_j \kappa_c/(\kappa_c+\kappa)^2$, which we also plot in figure \ref{Figure2}(b). It is maximal when $\kappa_j = \kappa_c+\kappa_i$, which happens at a voltage slightly above critical coupling.
	The expected efficiency is steadily around 87\% over the full working window $2\Delta-\hbar\omega< eV < 2\Delta$ of the detector (gray area).

\begin{figure*}[htbp]
		\begin{center}
			\includegraphics[width=17cm]{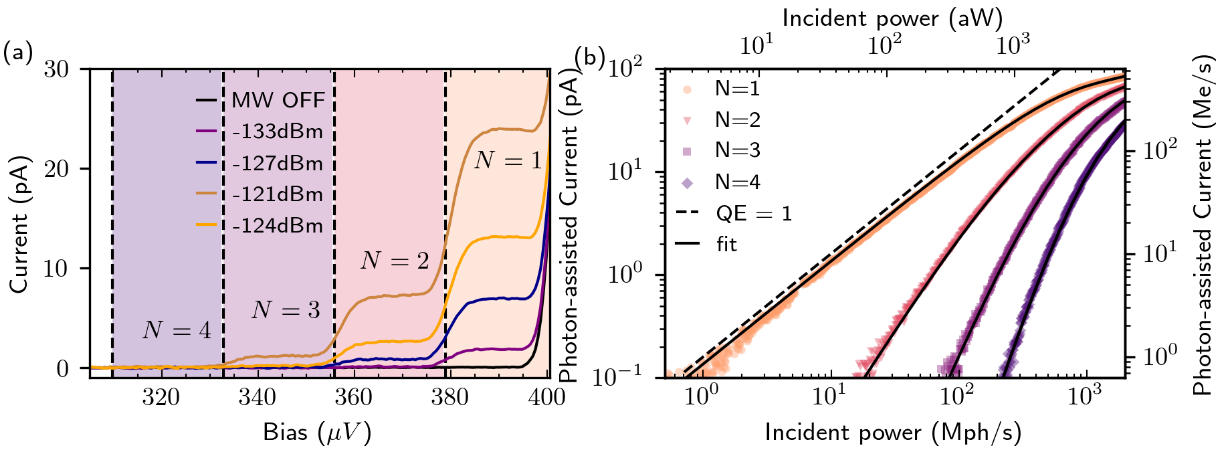}
		\end{center}
		\vspace{-0.8cm}
		\caption{\textbf{Photon-assisted current and quantum efficiency.} (a) 
			Voltage dependence of the subgap current for different microwave powers at frequency $\omega/2\pi =  \qty{5.525}{\GHz}$. At low power, a single step is observed when $2\Delta-\hbar\omega<eV<2\Delta$ corresponding to a PAT current where one photon gives one electron. At higher powers, multi-photon processes involving the tunneling of a single electron with $N$ absorbed photons occur when  $eV>2\Delta-N\hbar\omega$.
			(b) Power dependence of the $N=1,2,3,4$ PAT current steps and the corresponding theoretical prediction~\cite{SI}.}
		\label{Figure3}
		\vspace{-0.4cm}
	\end{figure*}

	\textbf{\textit{Photon-assisted tunneling current from coherent source---}}
	In order to verify that the losses induced by the junction indeed correspond to the conversion of photons into electrons, we measure the PAT current $\IPAT$ as a function of the incident microwave power.
	Figure \ref{Figure3}(a) shows the power dependence of the current-voltage characteristics when shining a microwave tone at the resonator frequency. As the power increases, a first subgap current step emerges that grows linearly with power.
	Additional steps appear at higher power that correspond to multiphoton absorption processes. Because of energy conservation, processes where $N$ photons are absorbed to photoassist the tunneling of one electron are allowed when $eV > 2\Delta - N \hbar \omega$. The coloured bands indicate the process with minimal $N$ allowed at each step.
	We measure the photocurrent at the center of the first four steps and plot their evolution with power in figure \ref{Figure3}(b). The plotted $\IPAT$ corresponds to the difference between the current with and without microwaves, not taking into account the dark current, which will be addressed later. At the $N=1$ step, the ratio between $I_{\rm PAT}/e$ and the photon flux, is the quantum efficiency $\chi$. 

	A precise direct measurement of $\chi$ is known to be difficult because of the unavoidable uncertainty on the value of the attenuation between the microwave source and the sample in the cryostat.
	In the limit $\nph \ll 1$, the PAT current at step $N$ is given by the population in the $N$ Fock state multiplied by the rate $|\mel{0}{D_\lambda}{N}|^2 \times I(V_m)/e$ where $V_m=(2\Delta + \hbar\omega/2)/e$. The population is given by $(\phi/\kappa_N)^N$ where $\kappa_N$ is an effective loss rate, which depends on $\kappa_c$, $\kappa$, $\lambda$ and $I(V)$ and can be calculated analytically~\cite{SI}.

	Fixing $\kappa_c$ to $2\pi \times \qty{75}{\MHz}$, $\lambda=0.79$, and using the independently measured resistance of the junction to estimate $I(V_m)=\qty{190}{\pA}$, the four $\IPAT$ curves can be fitted with the attenuation as a single free parameter, which is very constrained because each curve is proportional to the $N$-th power of the attenuation. Because the assumption $\nph \ll 1$ rapidly breaks down with increasing power, the fitting procedure is a bit more elaborate and relies on the numerical simulation of the Lindblad master equation describing the dynamics of the resonator mode instead of the analytical expressions~\cite{SI}. We obtain an attenuation of $\qty{-107.0\pm 0.3}{dB}$, corresponding to $\chi=83\%$. The resulting simulated $\IPAT$ curves are shown as solid lines in figure \ref{Figure3}(b). The good quality of the fit for all the steps at once with a single adjustable parameter confirms our good understanding of the PAT processes. We estimate the systematic uncertainty on the quantum efficiency to be on the order of 0.05.

	The deduced attenuation factor is only \qty{1.0}{\dB} above the estimated value that we obtain from an independent measurement of the overall transmission of the microwave lines of the cryostat and \qty{0.2}{\dB} above the one obtained by shot-noise calibration~\cite{SI}. At large power, the detector saturates when $\nph$ becomes on the order of unity and multiphoton processes cannot be neglected. More specifically the 1dB (3dB) compression points occur respectively for $P=-119.0\, (-114.2)$\,dBm or $\phi=0.34~(1.0)$\,Gph/s.
	%at which we estimate $\nph=2.3 \, (6.9)$.

	\textbf{\textit{Photon-assisted current from thermal population---}}
	\label{Calibration}
	As an independent test of our estimate of the quantum efficiency, we measure the conversion of the blackbody radiation coming from the matched \qty{50}{\ohm} load that is connected to the circulator port in direct view of the detector (see detailed circuit in~\cite{SI}). We suppose that the load is thermalized to the fridge temperature. In the limit $\nph \ll 1$, the expected equilibrium resonator population reads $\nph(T) = (\kappa_c+\kappa_i)/(\kappa_c+\kappa_i+\kappa_j)n_{\rm BE}(T)$ with the Bose-Einstein distribution $n_{\rm BE}(T) = 1/(\exp(\hbar \omega/k_B T) - 1)$. This assumes that the coupling and intrinsic loss channels both connect the resonator to a bath at temperature $T$ and that the junction behaves as a zero temperature bath. The corresponding thermally photon-assisted current is $e \kappa_j \nph$. Here, we measure the total current, which is given by $I_{\rm D} + e \kappa_j \nph$, where $I_{\rm D}$ is the dark current of our detector. 
	\begin{figure}[htbp]
		\vspace{-0.0cm}
		\begin{center}
			\includegraphics[width=8.5cm]{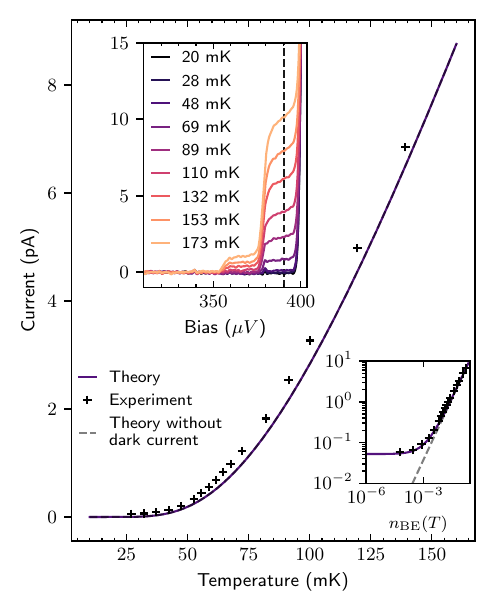}
		\end{center}
		\vspace{-0.5cm}
		\caption{\textbf{Thermal photon-assisted current and dark current.} 
			Temperature dependence of the current measured at the center of the $N=1$ photo-current step (dashed line in top inset). The full line is the expected current using the values of the loss rates fitted in figure \ref{Figure1} to which is added the measured dark current. The dashed line assumes zero dark current.
			Top inset:  Voltage dependence of the subgap current for different temperatures. Lower inset: same data as main in logarithmic scale vs $n_{\rm BE}(T)$. The saturation at low temperature is due to the dark current of our detector.}
		\label{Figure4}
		\vspace{-0.5cm}
	\end{figure}
	Figure \ref{Figure4} shows the measured current-voltage traces (inset) and the current at fixed bias voltage in the middle of the $N=1$ step as a function of temperature from 20 to \qty{150}{\milli\K}. The full line corresponds to the expected current including a dark current $I_{\rm D}=\qty{55}{\femto\A}$, while the dashed line shows the expected photocurrent $e \kappa_j \nph$ alone. Again, the agreement between the simple prediction using the loss rates obtained from the resonator spectroscopy and the measured current is good without any adjustable parameters. The small discrepancy is explained if we consider the sample temperature to be \qty{3}{\milli\K} above the one measured by the thermometer.
	The voltage dependence of the dark current (see~\cite{SI}) indicates that it is not due to a rounding of the $I(V)$ of the junction near the gap but rather to a non equilibrium population of the resonator mode which we estimate to $\nph=10^{-3}$, which corresponds to a temperature of \qty{40}{\milli\K}, close to the fridge temperature. We expect that a better shielding and filtering of the setup should reduce this value~\cite{Albertinale2021}.
	
	\textbf{\textit{Noise Equivalent Power---}}
	\label{sectionFigureMerit}
	In the data presented here, the current is measured using a differential voltage amplifier on a \qty{51.6}{\kilo\ohm} resistor in series with the junction. The equivalent current noise of this setup is  $120\,\rm{fA}/\sqrt{\rm{Hz}}$, which results in a Noise Equivalent Power (NEP) of $3.3\times 10^{-18} \, \rm{W}/\sqrt{\rm{Hz}}$. The setup could be modified in order to reduce the current noise measurement to $\delta I = 1\,\rm{fA}/\sqrt{\rm{Hz}}$ by using a HEMT voltage amplifier \cite{Jin2016} and increasing the value of the resistor in series with the junction, reducing the NEP to $3\times 10^{-20}\,\rm{W}/\sqrt{\rm{Hz}}$ on par with bolometric detection \cite{Kokkoniemi2020}.
	
	\textbf{\textit{Conclusion---}}
	We have realized an important proof of concept regarding continuous and efficient low energy microwave photon to electron conversion using a high impedance quantum circuit. The quantum efficiency of the process is unprecedentedly high (83\%) compared to the state of the art \cite{Khan2021} and is limited by the internal quality of the grAl resonator, which has been shown by others to be high \cite{Grunhaupt2018}. The relatively large intrinsic losses of the sample are of unknown origin. Other resonators fabricated with the same technique have shown significantly larger intrinsic quality factors (about two times). Quantum efficiency as large as 99\% should be within reach of this technique by increasing the intrinsic quality factor above $10^4$. We anticipate direct application to single microwave photon detection using quantum tunneling in a localized dot~\cite{Wong2017,Cornia2021,Khan2021,Haldar2024a,Haldar2024b}, in which the charge can be probed in real time~\cite{Haldar2024b} with a radio-frequency single electron transistor~\cite{Schoelkopf1998}. Given the charge sensitivity of state-of-the-art single electron transistor, single photons could be measured at a maximal rate of 100\,kph/s. This will require to lower the detection bandwidth ($\kappa_c$). The recent work in \cite{Albertinale2021} shows that dark count rates as low as \qty{100}{\Hz} can be reached in a narrow band detector. Compared to superconducting qubit based detectors, continuous operation is possible and does not require high frequency pump signals. Also, qubit based detectors, though very efficient, can only be used in the photon counting regime and are intrinsically limited to low photon fluxes ($< \sim 100$\,kph/s). Our detector can handle large fluxes and performs better than an ideal power detector behind a Josephson Parametric Amplifier with a noise temperature of one photon when the measurement bandwidth is larger than $\delta I^2/(\eta e)^2 \simeq \qty{50}{\MHz}$. Finally, the very same detector, here demonstrated around \qty{6}{\GHz}, could be used at frequencies as high as twice the superconducting gap ($\approx \qty{100}{\GHz}$) by tuning the resonator dimensions. This could be of interest to axion search \cite{Sikivie2021,Pankratov2022}.
	%		
	%	As an extension of this approach, the very same principle, here demonstrated around \qty{6}{\GHz}, can be applied for microwave frequencies as high as twice the superconducting gap($\approx 100$~GHz). At higher frequencies, other types of superconducting detectors, for instance kinetic inductance detectors, become efficient. The frequency could also be lowered, which will most likely result in a larger dark current. The quantum efficiency will remain high as long as the different steps in the photocurrent are well resolved.
	%
	%
	
	%	 We anticipate direct application in single microwave photon detection \cite{Lescanne2020,Inomata2016} using quantum tunneling in a localized dot\cite{Wong2017,Ghirri2020,Khan2021,Haldar2024a,Haldar2024b}. Compared to superconducting qubit based detectors, the principle of operation is continuous and does not require high frequency pump signals which limit the operation frequency to 40 GHz typically. The very same principle, here demonstrated around \qty{6}{\GHz}, can be applied for microwave frequencies as high as twice the superconducting gap ($\approx 100$~GHz) which is an important figure of merit regarding axion search \cite{Sikivie2021,Pankratov2022}.
	%
	%	At higher frequencies, other types of superconducting detectors, for instance kinetic inductance detectors, become efficient. 
	%	The frequency of operation can also be lowered, which will most likely result in a larger dark current. The quantum efficiency will remain high as long as the different steps in the photocurrent are well resolved. 
	%	

	\textbf{\textit{Acknowledgements---}}
	We thank R. Deblock, C. Quay Huei Li, A. Palacio-Morales, for stimulating discussions, C. Di Giorgio for her help in cabling the dilution fridge, L. Galvao-Tizei for TEM images.
	We acknowledge financial support from the ANR (ANR-21-CE47-0010 KIMIDET project and ANR-18-CE47-0003 BOCA project), from the France 2030 plan under the ANR-22-PETQ-0003 RobustSuperQ grant, from the Région Île-de-France in the framework of DIM SIRTEQ (Science et Ingénierie en Région Île-de-France pour les Technologies Quantiques) and the Laboratoire d’excellence Physique Atomes Lumière Matière (ANR-10-LABX-0039-PALM).

\appendix

\renewcommand{\thefigure}{S\arabic{figure}}
\renewcommand{\thetable}{S\arabic{table}}
\renewcommand{\theequation}{S.\arabic{equation}}
\setcounter{figure}{0}
\setcounter{table}{0}
\setcounter{equation}{0}

\section{Sample Fabrication and Experimental setup}
The sample shown in figure 1 of the main text is fabricated via a three steps process using standard e-beam and optical lithography techniques on an oxidized silicon wafer. We first write by e-beam lithography a narrow (\qty{720}{\nano\metre}) and long (\qty{180}{\micro\metre}) nanowire. Then we evaporate a thin (\qty{20}{\nano\metre}) film of granular aluminum.
In a second step we draw by e-beam lithography the design for a small ($140$~nm$\times120$~nm) Al-based superconducting tunnel junction that we evaporate using the Manhattan technique with three successive oxidation steps to obtain a large normal state resistance. In the last step we use optical lithography to draw and evaporate $100$~nm of aluminum to realize a \qty{50}{\ohm} microstrip transmission line and connect the tunnel junction and the grAl with a patch.

\begin{figure}[hbtp]
	\begin{center}
		\includegraphics[width=7cm]{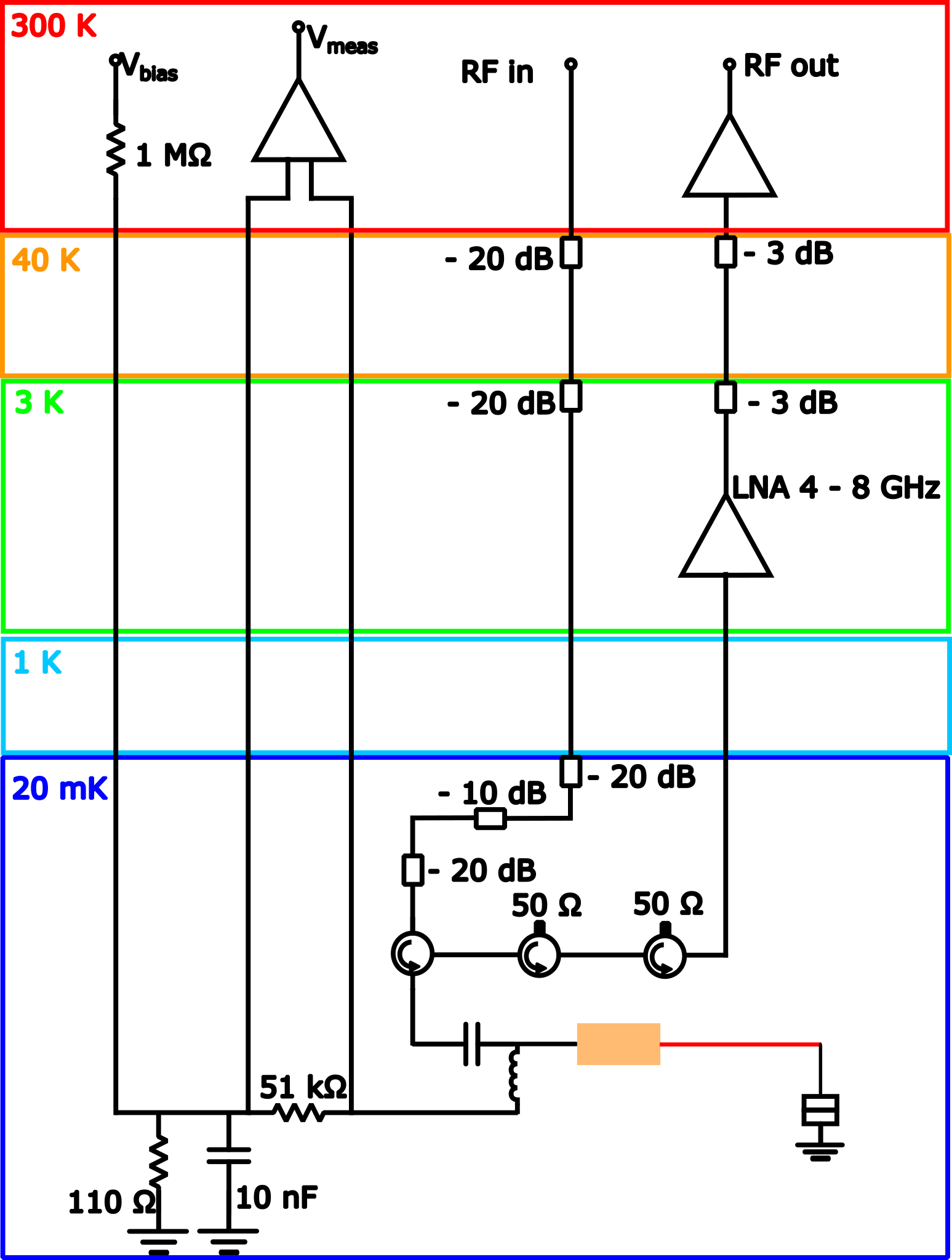}
	\end{center}
	\vspace{-0.3cm}
	\caption{Detailed experimental setup}
	\label{fridge}
	%\vspace{-0.3cm}
\end{figure}

The device is mounted in a copper box equipped with a microwave printed circuit board. The box is sealed and anchored to the mixing chamber plate of a dilution fridge with a base temperature of \SI{20}{\milli\kelvin}. The experimental setup is shown in figure  \ref{fridge}. In order to dc-bias the junction and do microwave measurements concurrently, the device is connected to the rest of the measurement circuit through a diplexer or a bias tee. The microwave measurements were carried out with a vector network analyzer. The excitation is delivered to the device through an attenuated line, and the reflected signal is directed to the output line using a series of circulators. The reflected signal is amplified first by a low temperature HEMT amplifier anchored at \SI{4}{\kelvin}, and then by two room temperature amplifiers. At the same time, the junction is voltage biased using a voltage divider and a filtering capacitor connected to the low frequency port of the diplexer/bias-tee at base temperature. Current through the junction is measured by the voltage drop across a \SI{51.6}{\kilo\ohm} resistor connected in series with the device. The voltage is amplified using a NF LI-75A differential preamplifier, prior to being further amplified and filtered by a Stanford Research SR560 amplifier. The gain of this measurement chain, as well as the measurement resistor has been calibrated in a separate cooldown. The input impedance of the preamplifier was measured independently to be \SI[separate-uncertainty = true]{94.8(2)}{\mega\ohm}.

\section{Resonator Characterization}\label{resonator}
The quarter wavelength granular aluminum (grAl) resonator is designed such that the fundamental resonance is close to \qty{6}{\giga\hertz} with a characteristic impedance of approximately \qty{5}{\kilo\ohm}. In order to characterize the resonator, we measure its resistance in the normal state as function of temperature during the cooldown. Just above the superconducting transition, we obtain $\SI[separate-uncertainty = true]{243(3)}{\kilo\ohm}$. Knowing the wire dimensions (\qty{180}{\micro\meter} long, and \qty{720}{\nano\meter} wide), we obtain the sheet resistance of the grAl layer to be $R_\square = \SI[separate-uncertainty = true]{970(20)}{\ohm/\sq}$, which corresponds to a resistivity of $\SI[separate-uncertainty = true]{1940(40)}{\micro\ohm\centi\metre}$. Using Mattis-Bardeen theory in the low temperature limit, the expected sheet inductance is $L_{\square} = \hbar R_{\square}/(\pi \Delta_{\rm grAl})$, where $\Delta_{\rm grAl} = \SI{330}{\micro\electronvolt}$ is the grAl superconducting gap, which we independently measure using tunneling spectroscopy. We obtain a sheet kinetic inductance $L_\square = \SI[separate-uncertainty = true]{620(10)}{\pico\henry/\sq}$. 

We simulate the electromagnetic properties of the resonator using a simple transfer matrix method. We input the sheet inductance determined above and the capacitance per unit length that we determine to be \SI{41.7}{\pico\farad\per\metre} from finite element simulations using the Sonnet Software. We calculate the resonant modes by supposing that the grAl resonator is terminated by a capacitor representing the junction. The first two modes are experimentally measured to resonate at $\omega/2\pi = \SI{5.52}{\giga\hertz}$, and $\omega_2/2\pi = \SI{17.8}{\giga\hertz}$. The junction capacitance is chosen such that these values are reproduced by the model. This capacitance is \SI{2.6}{\femto\farad}, which is compatible with the \qtyproduct{120x140}{\nano\metre} junction and the \qtyproduct{20x20}{\micro\metre} patching pad, which connects the junction to the resonator. Using the same parameters, we compute the resonant frequencies of the higher modes, as well as their impedance, and coupling parameter. The results are given in the table below:
\begin{table}[h!].
	\centering
	\begin{tabular}{|c |c |c |c|} 
		\hline
		Mode & $\omega/2\pi$ (\SI{}{\giga\hertz}) & $Z_c$ (\SI{}{\kilo\ohm}) & $\lambda$ \\ 
		\hline
		1 & 5.52 & 5.09 & 0.79 \\ 
		2 & 17.77 & 0.77 & 0.31 \\
		3 & 31.39 & 0.20 & 0.16 \\
		4 & 45.59 & 0.07 & 0.09 \\
		5 & 60.04 & 0.03 & 0.06 \\ 
		6 & 74.59 & 0.02 & 0.05 \\ 
		7 & 89.21 & 0.01 & 0.04 \\ 
		\hline
	\end{tabular}
\end{table}

\section{Junction Characterization}\label{junction}
The previous model also predicts a value for the coupling loss rate, which is $\kappa_c / 2\pi \approx \SI{78}{\mega\hertz}$. As explained in the main text, we need to match this rate to the junction loss rate (see equation 1 in the main text). A junction with a normal state resistance of $R_N \approx 2\Delta\lambda^2\rm{exp}(-\lambda^2)/(e^2 \kappa_c) = \qty{1.7}{\mega\ohm}$ should satisfy this condition, where the superconducting gap of aluminum is $\Delta = \qty{203}{\micro\electronvolt}$. 

In order to precisely estimate the photon-assisted tunneling rate for the fundamental mode, the effective resistance that enters the calculation of the $I(V)$ characteristics must be corrected for the dynamical Coulomb blockade factors due to the higher modes. We suppose that these modes remain in vacuum, and we renormalize the tunnel resistance of the junction compared to the value of \qty{1.53}{\mega\ohm} measured at high bias voltage as
\begin{equation}
	R_T =\prod_{n\neq 1}|\bra{0}\rm{exp}(i\lambda_n(a_n +a^\dagger_n))\ket{0}|^2\cross \qty{1.53}{\mega\ohm} = \rm{exp}(\lambda_2^2 + \lambda_3^2 + ...)\cross \qty{1.53}{\mega\ohm}.
\end{equation}
The resulting resistance $R_T = \qty{1.75}{\mega\ohm}$ is used in all our simulations. We then model the $I(V)$ characteristic of the junction through the standard formula
\begin{equation}
	I(V) = \frac{1}{eR_T}\int_{-\infty}^{\infty} n_L(E)n_R(E + eV)(f(E) - f(E+eV)) \, dE
\end{equation}
where $f(E)$ is the Fermi distribution, and $n_L$ and $n_R$ are the dimensionless quasiparticle density of states of the left and right junction electrodes
\begin{equation}
	n(E) = \Re\left(\frac{E + i\Gamma}{\sqrt{(E+i\Gamma)^2 - \Delta^2}}\right).
\end{equation}
The density of states is parameterized through the gap $\Delta$ and the Dynes parameter $\Gamma$. In our experiment, this parameter is too small to be reliably determined, therefore we consider $ \Gamma = \qty{0.01}{\micro\electronvolt}$, which has negligible influence on the results but eases the numerical convergence of the integral. We extract the gap value $\Delta = $ \SI{203}{\micro\electronvolt} from the measured current-voltage characteristic of the junction. The value of the current just above the gap at $eV=2\Delta + \hbar \omega/2$ is then \qty{190}{\pA}.

\section{Quantum Master Equation and Power Calibration}%\ref{calibration}
In order to describe the resonator coupled to the junction, we consider a quantum master equation as detailed in \cite{Esteve2018}. The coupling between the resonator mode and the junction results in an irreversible energy exchange, which is represented by quantum jumps, and in a Lamb shift of the resonator eigenstates. Quantum jump operators $A_l$, which describe the non-unitary evolution of the system, create (annihilate) $l$ ($-l$)  photons in the resonator for $l > 0$ ($l < 0$). The jump operators are given in Fock-state basis by \cite{Glauber}:
\begin{equation}
	\bra{n + l}A_l\ket{n} = \left(\frac{n!}{(n+l)!}\right)^{1/2}(i\lambda)^l \rm{exp}(-\lambda ^2/2)L_n^{(l)}(\lambda^2),\ l \geq 0
\end{equation}
where $L_n^{(l)}$ are the generalized Laguerre polynomials. Operators $A_l$ for $l < 0 $ are defined as $A_{- l} = (-1)^l A_l^\dagger$. Each quantum jump operator enters the master equation with a corresponding jump rate, which, to a good approximation, is proportional to $I(V- l\hw /e)$, where $I(V)$ is the bare junction characteristics (see previous section). The Hamiltonian part of the evolution is governed by
\begin{align}
	H &= H_D +  H_{\mathrm{LS}} \\
	H_D &= i\eta(a - a^\dagger) - \delta a^\dagger a \\
	H_{\mathrm{LS}} &= -\frac{1}{2e}\sum_{l = - \infty}^{\infty}I^{\mathrm{KK}}(V + l\hw /e)A_lA_l^\dagger
\end{align}
where $H_D$ describes the drive, with $\eta$ being proportional to the drive amplitude and $\delta$ the detuning. The rate $\eta$ is related to the incident photon flux $\phi$ by $\eta^2 = \phi \kappa_c$. The Lamb shift term $H_{\mathrm{LS}}$ accounts for the frequency shift induced by the junction and does not play an important role in the results presented here, the strength of the shift is proportional to the Kramers-Kronig transform $I^{\mathrm{KK}}(V)$ of the $I(V)$ characteristics. Except for the attenuation factor $A$, which reduces the flux $\phi$ by an unknown amount, the other parameters entering the master equation are fixed to their \emph{ab initio} values. The resonator loss $\kappa = \kappa_c + \kappa_i$ is taken from the mode spectroscopy (see figure 2 in the main text), the $I(V)$ characteristics of the junction is the one described in the previous section and the coupling parameter is set to $\lambda = 0.79$ as detailed above. 

We look for the steady state solution of the following master equation
\begin{equation} \label{meq}
	\frac{\mathrm{d}\rho}{\mathrm{d}t} = -i[H,\rho] +\frac{1}{e}\sum_{l = - \infty}^{\infty}I(V - l\hw /e)\mathcal{D}[A_l](\rho) + \kappa \mathcal{D}[a](\rho)
\end{equation}
where $\mathcal{D}[F](\rho) = F\rho F^\dagger - \frac{1}{2}\{F^\dagger F,\rho\}$, and $\rho$ is the reduced density matrix of the resonator. The last term accounts for the damping of the resonator with a loss rate $\kappa = \kappa_c + \kappa_i$ as detailed in the main text. We assume that the thermal population of the corresponding bath is negligible. From the steady state solution $\rho_s$, we obtain the expected photo-assisted current through
\begin{equation}
	I_{\mathrm{PAT}} = \sum_{l  \neq 0} I(V - l\hw /e)\ \Tr[A_l^\dagger A_l \rho_s]
\end{equation}

\begin{figure}[h]
	\begin{center}
		\includegraphics[width=8.9cm]{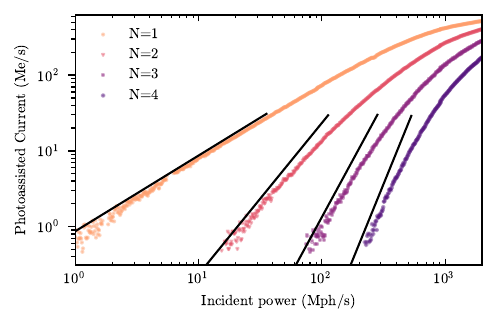}
	\end{center}
	\vspace{-0.3cm}
	\caption{Photo-assisted current as a function of pump power for bias voltages $eV = 2\Delta - (N-1/2)\hw, N = 1,...,4$ (same data as in figure 3 of the main text). Black solid lines are analytic predictions obtained by expanding the solution of the master equation at low power (equations (\ref{eq.Ipat1}) to (\ref{eq.Ipat4})).}
	\label{lpqme}
	%\vspace{-0.3cm}
\end{figure}

In order to calibrate the attenuation $A$, we measure the photo-assisted current as a function of microwave pump power for four different junction bias voltages $eV = 2\Delta - (N-1/2)\hw, N = 1, \ldots ,4$. In the low pumping limit where $n_{ph} \ll 1$, we can solve the master equation analytically at the corresponding voltages by truncating it to the lowest $N+1$ Fock states and by neglecting the different energy shifts. To lowest order in $\eta$, the photo-assisted current at each step is:
\begin{align}
	I_{\mathrm{PAT}}^{(1)} &= \frac{4\eta^2\kappa_j^{(1)}e}{(\kappa + \kappa_j^{(1)})^2} \label{eq.Ipat1} \\ 
	I_{\mathrm{PAT}}^{(2)} &= \frac{32\eta^4\kappa_j^{(2)}e}{\kappa^2(2\kappa + \kappa_j^{(2)})^2} \label{eq.Ipat2}  \\
	I_{\mathrm{PAT}}^{(3)} &= \frac{96\eta^6\kappa_j^{(3)}e}{\kappa^4(3\kappa + \kappa_j^{(3)})^2} \label{eq.Ipat3}  \\
	I_{\mathrm{PAT}}^{(4)} &= \frac{512\eta^8\kappa_j^{(4)}e}{3\kappa^6(4\kappa + \kappa_j^{(4)})^2} \label{eq.Ipat4}  \\
	\textrm{with}\ \kappa_j^{(n)} &= \frac{\rm{exp}(-\lambda^2)\lambda^{2n}}{n!}\frac{I(2\Delta/e + \hw/2e)}{e}
\end{align}
where $I_{\mathrm{PAT}}^{(N)}$ stands for the photo-assisted current at $eV = 2\Delta - (N-1/2)\hw$. As $\eta^2 \propto A$, fitting these four currents is equivalent to simultaneously fitting $A, A^2, A^3$ and $A^4$, which allows us to precisely calibrate $A$. Figure \ref{lpqme} shows the predictions of these simple analytical power law formula together with the data. As can be seen, the validity domain of the analytical formula is too restricted, which is why we rather use the numerical solution of the master equation \cite{qt2} in order to obtain a more precise fit as shown in the figure 3 of the main text. From this method, the estimated attenuation is $A=$\SI{107}{\decibel}.

\subsection*{Other Methods of Calibration}
We have verified that two other calibration methods give consistent results with the method detailed in the previous section. We first estimate the attenuation of every microwave component and cable and obtain a total attenuation of \SI{106}{\decibel}. This method relies on many assumptions and cannot be considered as reliable. We also consider the shot-noise emission of the junction at a large bias voltage as a calibrated source that allows us to know the gain of the amplification chain, from which we obtain the attenuation by measuring the reflected power outside the resonance. 

At sufficiently high bias compared to the gap, where the current $I$ through the junction can be approximated as being proportional to $V$, the noise power emitted by the junction in a bandwidth $BW$ centered at the resonant frequency of the resonator is \cite{Menard2022}
\begin{equation}\label{snfit}
	P = \frac{4  \lambda^2 \kappa_c \hbar \omega I}{e(\kappa + \omega \lambda^2 R_N/\pi R_K)^2} \cross BW
\end{equation}
where $R_N$ is the measured resistance of \qty{1.53}{\mega\ohm}. We measure the emitted power in a \qty{5}{\mega\hertz} bandwidth around resonance between \qty{1.3}{\milli\volt} and \qty{1.8}{\milli\volt}, which increases linearly with $V$ as expected. By comparing the slope to equation \ref{snfit}, we obtain the gain of our measurement chain, from which we deduce an attenuation of $\SI[separate-uncertainty]{107(1)}{dB}$. 

\section{Quantum Efficiency Error Estimation}
\begin{figure*}[htbp]
	\begin{center}
		\includegraphics[width=17.8cm]{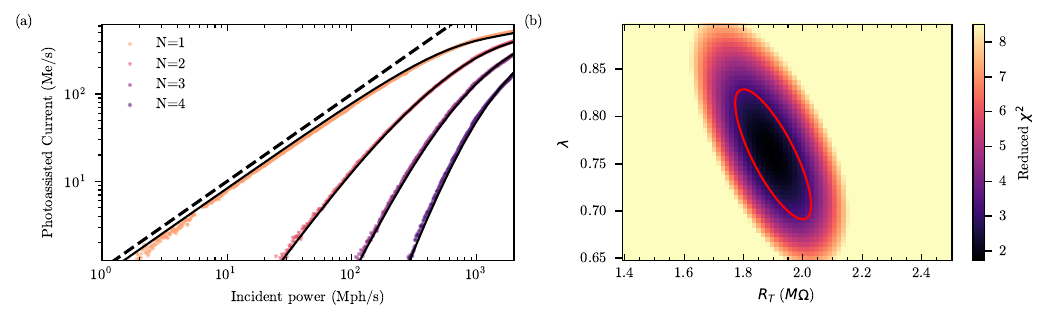}
	\vspace{-0.3cm}
	\caption{(a) Fit of the photon-assisted current measured at bias voltages  $eV = 2\Delta - (N-1/2)\hw, N = 1,...,4$ (same data as in figure 3 of the main text), but $\lambda$ and $R_{T}$ are now free fit parameters. The dashed line represents ideal quantum efficiency. (b) Reduced $\chi_r^2$  as a function of $\lambda$, and $R_T$. For each value of these two parameters, we obtain a fitted attenuation and a reduced $\chi_r^2$. By considering the region where $\chi_r^2$ remains below two times its minimal value (red ellipse), we obtain an estimation of the error on the attenuation from the standard deviation of the different attenuation results.}
	\label{qmefree}
	\end{center}
	%\vspace{-0.3cm}
\end{figure*}

Once the attenuation of the incoming photon flux is known, the quantum efficiency is directly obtained from the linear increase of the photon-assisted current as a function of flux. As stated in the main text, we obtain a quantum efficiency of \SI{0.83}{} with negligible statistical error given the negligible noise on the current as seen in figure 3 of the main text. The main source of error is the systematic error that enters the calibration of the attenuation. In particular, we have to know the effective resistance $R_T$, which includes the effect of dynamical Coulomb blockade of all the higher modes that we cannot characterize experimentally. In order to estimate the effect of this error, we let $R_T$ and $\lambda$ be free fit parameters in the calibration procedure. We then obtain $\lambda = 0.76, R_{T} = \SI{1.90}{\mega\ohm}$ and $A =\SI{106.7}{\decibel}$, which results in a quantum efficiency of 0.79. With these extra fit parameters, the reduced $\chi_r^2$ is below 2, which indicates that it can be used to estimate the error bar on the fit parameters. By considering the region where the reduced $\chi_r^2$ is doubled compared to the optimal point as our uncertainty region (see figure \ref{qmefree}), we obtain a 1$\sigma$ error bar of 0.05 on the quantum efficiency, which is the value of the error stated in the main text. 

\section{Dark Current}
\begin{figure}[htbp]
	\begin{center}
		\includegraphics[width=8.9cm]{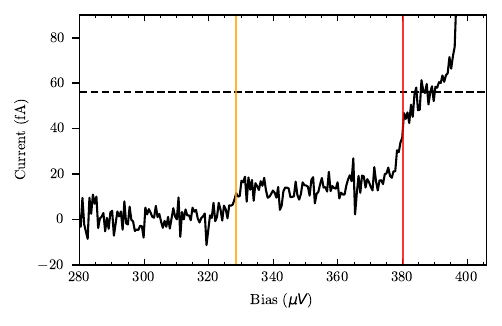}
	\end{center}
	\vspace{-0.3cm}
	\caption{$I(V)$ characteristic of the junction at \SI{20}{\milli\kelvin} in the absence of any microwave drive. Two steps are still visible, corresponding to the 17.8\,GHz mode (orange line), and the fundamental mode (red line).}
	\label{dark}
	%\vspace{-0.3cm}
\end{figure}

Dark current is a key figure of merit for a photon detector. In order to obtain its value, we measure the current through the junction close to the gap in absence of any microwave drive. This measurement is shown in figure \ref{dark}.
We observe two steps, which indicate the presence of a residual photon population in the first two modes. The voltage width of the steps (see figure \ref{dark}) matches the fundamental resonator mode at \SI{5.525}{\giga\hertz}, and the one of the next mode at \SI{17.8}{\giga\hertz}. In the working window of our detector ($2\Delta - \hw < eV < 2\Delta$), the dark current is approximately $\qty{55}{\femto\ampere}$, which corresponds to a population of \SI{1.5e-3} in the fundamental mode as can be seen also in figure 4 of the main text.

\end{document}